\pdfoutput=1

\documentclass[sigconf,nonacm]{acmart}

\usepackage[hang,flushmargin]{footmisc}
\usepackage{booktabs} %
\usepackage{subcaption} %
\usepackage[normalem]{ulem}
\usepackage[export]{adjustbox}
\usepackage{enumitem} %
\usepackage{multicol}
\usepackage{setspace}
\usepackage[linesnumbered,boxed,lined,ruled,vlined,noresetcount]{algorithm2e}
\usepackage{mathtools} %
\usepackage{cleveref}
\usepackage{multirow}
\usepackage{array}
\usepackage{balance}
\usepackage{rotating}
\usepackage{outlines} %
\usepackage{bm} %
\usepackage{latexsym} %
\usepackage{xspace} %
\usepackage{color}
\usepackage{makecell} %
\usepackage{amsthm} %
\usepackage[zerostyle=d]{newtxtt} %
\usepackage[T1]{fontenc}
\usepackage{xtab,afterpage}
\usepackage{pifont} %
\usepackage{listings}
\usepackage{tabularx}
\usepackage{gnuplottex}
\usepackage{xcolor}
\usepackage{algorithmic}
\usepackage{graphicx}
\usepackage{breqn}
\usepackage{ragged2e}

\let\oldnl\nl%
\newcommand{\nonl}{\renewcommand{\nl}{\let\nl\oldnl}}%

\definecolor{commentcolor}{RGB}{107,142,35}

\definecolor{codegreen}{rgb}{0,0.6,0}
\definecolor{codegray}{rgb}{0.5,0.5,0.5}
\definecolor{codepurple}{rgb}{0.58,0,0.82}
\definecolor{backcolour}{rgb}{0.95,0.95,0.92}
\definecolor{ballblue}{rgb}{0.13, 0.67, 0.8}
\definecolor{babyblue}{rgb}{0.54, 0.81, 0.94}
\definecolor{bleudefrance}{rgb}{0.19, 0.55, 0.91}
\definecolor{awesome}{rgb}{1.0, 0.13, 0.32}
\definecolor{brilliantrose}{rgb}{1.0, 0.33, 0.64}
\definecolor{brightpink}{rgb}{1.0, 0.0, 0.5}
\theoremstyle{definition}

\theoremstyle{example}

\newcommand{\sql}[1]{{\color{awesome}{#1}}}

\newcommand{\uexp}{U-expression\xspace}
\newcommand{\uexps}{U-expressions\xspace}

\newcommand{\values}{\textsf{VALUES}\xspace}
\newcommand{\sel}{\textsf{{SELECTION}}\xspace}

\newcommand{\join}{\textsf{JOIN}\xspace}
\newcommand{\innerj}{\textsf{INNER JOIN}\xspace}
\newcommand{\crossj}{\textsf{CROSS JOIN}\xspace}
\newcommand{\fullj}{\textsf{FULL JOIN}\xspace}
\newcommand{\outerj}{\textsf{OUTER JOIN}\xspace}

\newcommand{\order}{\textsf{ORDER BY}\xspace}
\newcommand{\limit}{\textsf{LIMIT}\xspace}
\newcommand{\off}{\textsf{OFFSET}\xspace}

\newcommand{\union}{\textsf{UNION}\xspace}
\newcommand{\unionall}{\textsf{UNION ALL}\xspace}
\newcommand{\intersect}{\textsf{INTERSECT}\xspace}

\newcommand{\excepts}{\textsf{EXCEPT}\xspace}

\newcommand{\proj}{\textsf{{PROJECTION}}\xspace}
\newcommand{\case}{\textsf{CASE WHEN}\xspace}
\newcommand{\exist}{\textsf{EXISTS}\xspace}

\newcommand{\sqlover}{\textsf{OVER}\xspace}
\newcommand{\partitionby}{\textsf{PARTITION BY}\xspace}

\newcommand{\spark}{Spark SQL\xspace}

\newcommand{\sys}{\textsc{SQLution}\xspace}
\newcommand{\polygon}{\textsc{Polygon}\xspace}
\newcommand{\verieql}{\textsc{VeriEQL}\xspace}
\newcommand{\sqlsolver}{\textsc{SQLSolver}\xspace}
\newcommand{\wetune}{\textsc{WeTune}\xspace}

\newcommand{\genrewrite}{\textsc{GenRewrite}\xspace}

\newcommand{\udp}{\textsc{UDP}\xspace}

\newcommand{\cubes}{\textsc{Cubes}\xspace}

\newcommand{\revised}[1]{{#1}}

\usepackage[skins]{tcolorbox} %
\tcbuselibrary{breakable}
\newtcolorbox{highlight}[1]{  %
  enhanced,
  breakable,
  colback=white,
  colframe=green,
  coltitle=white,
  colbacktitle=green,
  fonttitle=\bfseries,
  title={#1},
  sharp corners,
  top=0mm,bottom=0mm,left=0mm,right=0mm,
  boxrule=1pt,
  before upper={\parindent10pt}  %
}

\newcommand\vldbpagestyle{standardpagestyle}

\begin{document}

\title{Bound to be Complete: Efficient Verification for SQL Equivalence via U-Expressions}

\author{Haoran Ding$^{1,2,*}$, Zhaoguo Wang$^{1,\dagger}$, Zuming Jiang$^3$, Haibo Chen$^1$, Jinyang Li$^4$, Ruzica Piskac$^5$}
\affiliation{%
  \institution{$^1$Institute of Parallel and Distributed Systems, Shanghai Jiao Tong University\quad   $^2$MIT CSAIL \\
  $^3$The University of Hong Kong\quad $^4$New York University\quad $^5$Yale University \\
  }
}

\begin{abstract}

Automated verification of SQL query equivalence is a fundamental problem in database research.
Recent works developed verifiers with guaranteed completeness, which can generate counterexamples to disprove non-equivalent queries and are useful in various scenarios.
However, they have to solve complex formulas and can fail to produce counterexamples within a reasonable time budget.

This paper presents a new complete verifier called \sys, which not only greatly reduces the verification time, but also supports more SQL features.
The key insight is to represent SQL semantics using U-expressions, a functional algebra that can model SQL with significantly simpler formulas than prior works, thereby leading to more efficient SMT solving.
To turn this insight into a practical verifier, \sys further tackles two technical challenges associated with modeling SQL using U-expressions, namely, the lack of tuple ordering and the presence of unbounded summations.
In particular, to support tuple ordering, \sys proposes \emph{order semantics encoding} which combines U-expressions with \revised{first-order logic (FOL)}.  To handle unbounded summations, \sys introduces \emph{unbounded summation expansion} to transform them into bounded summations, which can be directly encoded into FOL formulas.
Compared with state-of-the-art verifiers, \sys supports all 271 non-equivalent query pairs derived from Spider and reduces verification time by up to 98.6\%.

\end{abstract}
 
\maketitle

\begingroup
\renewcommand\thefootnote{*}\footnotetext{This research was completed when Haoran Ding was a PhD student in Shanghai Jiao Tong University. He is currently a postdoctoral associate of MIT.}
\renewcommand\thefootnote{$\dagger$}\footnotetext{Corresponding author.}
\endgroup

\pagestyle{\vldbpagestyle}

\vspace{-8pt}
\section{Introduction}
\label{sec:intro}

Automated verification of SQL query equivalence is one of the fundamental problems in database theory~\cite{abiteboul1995foundations}.
Two queries are considered equivalent if they yield identical results for all possible database instances.
Developing a verifier for such equivalence is crucial across a wide range of scenarios.
\revised{
For example, such verifiers can help automatically discover rewrite rules~\cite{wang2022wtune,bai2023querybooster,jiang2025vldb,liu2024,chu2017hottsql}, which transform a given query into another equivalent but faster query.
Verifiers can also help detect database bugs~\cite{mang2025llm,he2024verieql}.
Some bugs cause query optimizers to rewrite queries into non-equivalent ones~\cite{he2024verieql}.
}
However, the SQL equivalence problem is undecidable~\cite{chu2018udp,Chandra1977optimal}, so no verifier can ensure both soundness and completeness.

Recently, there has been a rising interest in building complete verifiers~\cite{chu2017cosette,he2024verieql,zhao2025polygon} that automatically search for such counterexamples.
This is useful in several important scenarios, such as database bug detection~\cite{he2024verieql}, example-based query synthesis~\cite{rajeev2013cegis,dong2023slabcity,zhang2021oopsla,brancas2025cubes,brancas2024towards,wang2017sigmod,wang2017pldi}, query disambiguation~\cite{yang2025sqldriller}, and SQL grading~\cite{chandra2019icde,chandra2015data}.
In contrast, sound verifiers~\cite{sigmod2024sqlsolver, wang2022wtune, zhou2019equitas, zhou2022spes, wang2024qed, chu2017cosette, chu2018udp} usually cannot provide counterexamples because they are inherently incomplete and may falsely treat equivalent queries as non-equivalent.

State-of-the-art complete verifiers~\cite{he2024verieql,zhao2025polygon} model SQL tables as lists and propose list-based bounded symbolic reasoning.
In particular, the verifier sets a bound $b$ on table size, which allows it to represent each input table using a list of $b$ symbolic tuples. By symbolically evaluating a SQL query on these symbolic lists, the verifier can systematically encode the query into a first-order logic (FOL) formula, thus enabling the use of an SMT solver~\cite{Haniel2022cvc5,de2008z3,Clark2011cvc4} to produce concrete assignments to symbolic tuples as a counterexample that demonstrates non-equivalence.

Despite its promise, the existing approach for complete verification is inefficient: the list-based encoding yields complex formulas, resulting in a long running-time for SMT solvers.
For example, for a pair of queries with a total of 29 AST nodes from the Spider benchmark, the resulting formula has 2,916 AST nodes even if the table size bound is only 1.
When the bound increases to 3, the number of AST nodes in the formula increases to 18,001.
The inefficiency is due to two fundamental causes. First, symbolic execution leads to complex encoding.  In addition to modeling the input tables and final query result, existing verifiers also generate a symbolic list for the intermediate result of each SQL operator, which significantly increases the number of symbolic variables.
Moreover, for each tuple in the query result, the verifier generates a separate formula describing how that tuple is derived from the inputs.
The overall encoding is then expressed as the conjunction of all such formulas, resulting in expressions with numerous logical and arithmetic operators.
Second, aligning list semantics with SQL's bag semantics also results in complex encoding.
Under SQL's bag semantics, tuples in a table do not have a fixed order, and duplicates are allowed.
Consequently, when comparing query results under bag semantics, existing verifiers need to consider all possible permutations of tuples in the lists, producing large and complex formulas that challenge SMT solvers.

To overcome these limitations, our key insight is to encode query equivalence in a lightweight way, which is based on an algebra called U-expression~\cite{chu2018udp}.
Based on the key insight, we develop a new complete verifier called \sys.
The existing sound verifier \sqlsolver~\cite{sigmod2024sqlsolver} demonstrates that U-expressions can model a broad range of SQL features, including features unsupported by prior works~\cite{he2024verieql,zhao2025polygon}.
A U-expression formulates a SQL query as a function $f(t)$, which returns the multiplicity of any tuple $t$ in the query result.
This formulation compactly captures query semantics and enables equivalence reasoning using significantly fewer variables and simpler formulas.
Because the multiplicity of any tuple can be computed directly, \sys avoids generating separate formulas for each output tuple.
Moreover, query equivalence can be expressed in a particularly elegant form: two queries are equivalent if and only if their corresponding U-expressions yield the same integer value for any input tuple.
Formally, for U-expressions $f_1(t)$ and $f_2(t)$, equivalence checking reduces to proving the unsatisfiability of the formula $\exists t. f_1(t) \neq f_2(t)$.
Unlike comparing two symbolic lists, comparing two integer values is an encoding that is far simpler and more efficient to solve.

\revised{
Although \sqlsolver~\cite{sigmod2024sqlsolver} already uses U-expressions for sound verification, its algorithm to check the equivalence of U-expressions is inherently incomplete and cannot generate counterexamples for non-equivalent queries.
It is non-trivial to extend \sqlsolver to a complete verifier because query equivalence verification is undecidable.
Realizing a complete automated verifier based on U-expressions therefore requires solving two challenges that cannot be handled by \sqlsolver: reasoning about ordered bag semantics and encoding unbounded summations in U-expressions.
}
The first arises from the semantics of U-expression itself: by definition, U-expression captures only bag semantics and therefore cannot directly model queries with ordered bag semantics.
Those queries contain clauses such as \order, \limit, or \off~\cite{orderby}, for which the order among tuples in query results is significant.
Because U-expressions reason about the multiplicity of individual tuples rather than their relative order, they provide no native mechanism for expressing ordering among tuples.
\revised{
\sqlsolver supports \order through a syntax-matching algorithm, which cannot ensure completeness.
Specifically, it requires that:
1) The two queries contain the same number of \order clauses.
2) Each \order clause in one query matches another \order clause in the other query in terms of ordering directions and attributes.
3) The subqueries of each pair of matched \order clauses are equivalent.
This procedure is incomplete because semantically equivalent queries need not have matching syntactic structures.
Failure to find such a match therefore does not establish non-equivalence or yield a counterexample.
Because \sys must reason completely about ordered semantics, it cannot reuse this syntax-matching algorithm.
}

To overcome this challenge, we introduce \revised{an \emph{order semantics encoding}} technique for handling \order, \limit, and \off, \revised{which reasons about query equivalence by encoding the semantics rather than syntactic comparison, thereby ensuring completeness.}
Specifically, it systematically rewrites SQL queries to eliminate \limit and \off while preserving query semantics.
\order cannot be removed in general, because ordered bag semantics cannot be reduced to pure bag semantics.
Fortunately, we observe that while U-expressions can represent the multiplicity of only a single tuple in the query result, we can combine them with first-order logic formulas that explicitly encode the ordering relations between tuples.
Thus, we encode the existence of tuples $t_1$ and $t_2$ in the query result with two U-expressions. Then, we encode that $t_1$ precedes $t_2$ in one query result but the ordering is opposite in another query result based on first-order logic.
If the above constraint is satisfiable, queries are non-equivalent, and we will use SMT solvers to produce a counterexample.
For brevity, ordered bag semantics will be abbreviated as order semantics if the context is clear.

The second challenge is that there is no decision procedure to reason about U-expressions while ensuring completeness.
The main difficulty stems from unbounded summations ($\sum$) in U-expressions, which are hard to be directly encoded in first-order logic.
\revised{
In an unbounded summation $\sum_{t}e(t)$, the variable $t$ ranges over all possible tuples rather than a statically bounded set, which means the summation domain is unbounded.
}
Such summations are necessary to model SQL constructs such as \proj and aggregate functions.
However, a FOL formula cannot directly express the sum of unbounded expressions.
\revised{
\sqlsolver proposes an approximation-based method to encode U-expressions with unbounded summations, which cannot ensure completeness.
Specifically, the encoding result is an over-approximation formula such that if the formula is unsatisfiable, the original U-expressions are treated as equivalent.
However, if the formula is satisfiable, the original U-expressions may be either equivalent or non-equivalent.
Thus, \sqlsolver's method cannot determine whether two queries are non-equivalent and cannot produce a counterexample.
To ensure completeness, \sys cannot reuse \sqlsolver's encoding algorithm.
}

To address unbounded summations with a completeness guarantee, we introduce a new verification algorithm for U-expressions that extends the classical bounded verification idea~\cite{inverso2021bound,Baumann2021bound,Nagar2020bound,Kokologiannakis2020hmc,Kokologiannakis2019mc,Kokologiannakis2019lock,Kokologiannakis2017concurrency}.
By assuming that each table contains a bounded number of distinct tuples, we observe that every unbounded summation becomes equivalent to a bounded one, which can be directly encoded into FOL formulas.
Intuitively, we find that only a finite set of tuples can yield non-zero $e(t)$.
So enumerating all possible tuples in the unbounded summation domain is not really necessary.
Based on the observation, we propose \emph{unbounded summation expansion} to precisely construct a bounded domain for each unbounded summation, which covers each tuple $t$ that can make $e(t)$ non-zero.
Then, an unbounded summation can be expanded into an equivalent bounded summation.

We evaluated \sys on 997 SQL query pairs derived from Spider~\cite{yu2018spider}, a widely used benchmark for LLM-based query synthesis~\cite{Pourreza2023dinsql,li2024dawn,li2023aaai}.
\sys successfully handles all test cases, including 271 non-equivalent pairs.
In contrast, the state-of-the-art complete verifier \polygon supports only 716 cases, of which 162 are non-equivalent.
For non-equivalent queries disproved by both verifiers, \sys achieves substantial performance gains: it reduces verification time by over 50\% in 92.9\% of cases and by over 80\% in 43.6\% of cases, with a maximum time reduction of 98.6\% (from 802 ms to 11 ms).

\section{Background and Motivation}
\label{sec:back}

\subsection{Sound verifiers vs. Complete verifiers}
\label{subsec:back:verifier}

\revised{
The SQL query equivalence problem is undecidable~\cite{chu2018udp,Chandra1977optimal}, which means it is impossible for a verifier to ensure both \emph{soundness} and \emph{completeness}.
A sound verifier, such as \sqlsolver, guarantees that every query pair it proves equivalent is indeed equivalent over all database instances.
However, some queries may fail to pass the verification even if they are equivalent due to the inherent incompleteness of sound verifiers.
Thus, sound verifiers generally cannot produce a counterexample when they fail to prove equivalence.
In contrast to sound verifiers, this paper targets complete verifiers, which guarantee that queries that fail verification are indeed non-equivalent.
Complete verifiers produce a counterexample to demonstrate non-equivalence, i.e., a set of concrete database tables on which the two queries yield different results.
Due to undecidability, complete verifiers and sound verifiers usually adopt rather different algorithms.
It is intractable to extend a sound verifier, such as \sqlsolver, to a complete one.
To generate a counterexample to demonstrate the non-equivalence, complete verifiers typically use a bounded verification approach.
Given a bound $b$, a complete verifier searches for input tables such that each table size is within the bound and two queries produce different results.
If such a counterexample exists, the verifier demonstrates that the queries are non-equivalent.
Otherwise, the verifier concludes that the queries are equivalent when each table size is within the bound.
Increasing the bound $b$ allows the verifier to consider larger tables.
}

\subsection{Application of SQL Equivalence Verifiers}
\label{subsec:back:app}

\revised{
SQL equivalence verifiers are useful in various scenarios.
For example, sound verifiers can help automatically discover new rewrite rules~\cite{wang2022wtune,jiang2025vldb}.
Query optimizers typically apply rewrite rules to transform a query into another equivalent but faster one.
Existing works~\cite{wang2022wtune,jiang2025vldb} discover new rules automatically by first enumerating possible rewrite rules and using verifiers to identify ones that can ensure query equivalence before and after rewriting.
Complete verifiers are also useful in multiple scenarios.
First, they can help detect and understand bugs in database optimizers~\cite{he2024verieql,zhao2025polygon,mang2025llm}.
Although queries are equivalent by construction, the implementation of optimizers may have bugs~\cite{mysqlbug104918,duckdbissue11294,spark58383}.
Prior works~\cite{he2024verieql} have found such bugs in Apache Calcite and MySQL.
The counterexamples generated by complete verifiers can help developers understand and locate the bugs.
Second, the counterexamples generated by complete verifiers can be used to guide LLMs in rewriting SQL queries.
For example, \genrewrite~\cite{liu2026genrewrite} proposes a counterexample-guided approach that leverages an LLM to rewrite a given SQL query into an equivalent but faster query.
When the rewritten query is not equivalent to the original one, \genrewrite uses counterexamples to guide LLMs in correcting the rewrite.
Third, complete verifiers can help assess LLMs' capabilities in text-to-SQL tasks.
Text-to-SQL tasks require an LLM to translate natural-language questions into SQL queries.
Popular text-to-SQL benchmarks (e.g., Spider) can use complete verifiers to check whether generated SQL queries are equivalent to the reference SQL queries.
Effectively applying complete verifiers in these scenarios is outside the scope of this paper and remains an interesting topic for future work.
}

\subsection{List-based Verification for SQL Equivalence}
\label{subsec:back:priorwork}

\begin{figure}[t]
\begin{lstlisting}[language=SQL,breaklines=true,escapeinside={(*}{*)}]
create table hotel(id int, star int, price int);
(*$Q_1$*): (select * from hotel where star < 4) 
    union all
    (select * from hotel where price < 500)
(*$Q_2$*): select * from hotel where star < 4
\end{lstlisting}
\normalsize
\caption{An example of two non-equivalent SQL queries.}
\label{fig:motivating-example}
\end{figure}

We illustrate how a list-based complete SQL equivalence verifier operates on the example in \Cref{fig:motivating-example}.
\revised{In this example, the first query uses \unionall to concatenate the hotels whose star rating is below 4 with those whose price is below 500, duplicating tuples that satisfy both conditions. The second query returns only hotels whose star rating is below 4. Thus, the two queries are not equivalent.}
We begin with a bound of 1.
The list-based approach proceeds as follows.
The verifier first constructs symbolic lists to represent the input tables and query outputs.
Under a bound of 1, the symbolic list for the table {\tt{hotel}} contains at most one symbolic tuple:
$l_0 \coloneqq [(del_0, id_0, star_0, price_0)]$.
Here, the boolean variable $del_0$ indicates tuple deletion (\emph{i.e.}, the tuple does not exist if and only if $del_0$ is true), while the integer variables $id_0$, $star_0$, and $price_0$ correspond to the tuple attributes.
Given this bound, the result of $Q_1$ contains at most two tuples, and that of $Q_2$ at most one.
Accordingly, the symbolic lists for the query results are defined as
$l_1 \coloneqq [(del_1, id_1, star_1, price_1), (del_2, id_2, star_2, price_2)]$ for $Q_1$,
and
$l_2 \coloneqq [(del_3, id_3, star_3, price_3)]$ for $Q_2$,
where $del_1$, $del_2$, and $del_3$ denote the existence of the corresponding tuples.

Next, the semantics of $Q_1$ is encoded as $F_1$.
\begin{equation*}
\begin{aligned}
F_1 \coloneqq & \ (del_1 \leftrightarrow (del_0 \lor \neg (star_0 < 4)))\\
\land & \ (id_1=id_0) \land (star_1=star_0) \land (price_1 = price_0) \\
\land& \  (del_2 \leftrightarrow (del_0 \lor \neg (price_0 < 500)))\\
\land & \ (id_2=id_0) \land (star_2=star_0) \land (price_2 = price_0) \\
\end{aligned}
\end{equation*}
The resulting formula is a conjunction of two components.
The first component encodes the semantics of the first sub-query: when the tuple in table {\tt{hotel}} exists and satisfies the predicate $star < 4$, the corresponding output tuple in $Q_1$ 
must exist, and it is the output of the first sub-query.
The second line encodes the semantics of the second sub-query in the same manner.
Similarly, the semantics of $Q_2$ is encoded as $F_2$.
\begin{equation*}
\begin{aligned}
F_2 \coloneqq & \ (del_3 \leftrightarrow (del_0 \lor \neg (star_0 < 4)))\\
\land & \ (id_3=id_0) \land (star_3=star_0) \land (price_3 = price_0) \\
\end{aligned}
\end{equation*}

Finally, the equality of two lists $l_1$ and $l_2$ is encoded as $F_3$:
\begin{equation*}
\begin{aligned}
F_3 \coloneqq & \ (\sum_{t_i\in l_1}int(t_i.del) = \sum_{t_j\in l_2}int(t_j.del)) \\
\land & \bigwedge_{t\in l_1}\Bigl(\sum_{t_i\in l_1}int((t_i = t) \land \neg t_i.del \land \neg t.del) \\
& \quad = \sum_{t_j\in l_2}int((t_j = t) \land \neg t_j.del \land \neg t.del)\Bigr) \\
\end{aligned}
\end{equation*}
For simplicity, we use $int(p)$ to represent $ite(p, 1, 0)$.
If the formula $p$ is true, then $int(p)$ returns 1.
Otherwise, it returns 0.
The term $t.a$ represents the symbolic attribute of the symbolic tuple $t$.
$F_3$ consists of two parts.
The first part $\sum_{t_i\in l_1}int(t_i.del) = \sum_{t_j\in l_2}int(t_j.del)$ represents that the query results of $Q_1$ and $Q_2$ should have the same number of tuples.
The second part means that for each tuple $t$ in the list $l_1$, its multiplicity in the query result of $Q_1$ and $Q_2$ should be the same.
The multiplicity of $t$ in the result of $Q_1$ is calculated by $\sum_{t_i\in l_1}int((t_i = t) \land \neg t_i.del \land \neg t.del)$.
The multiplicity of $t$ in the result of $Q_2$ is expressed similarly.

In the final step,
the list-based verifier invokes an SMT solver to check the satisfiability of $F_1 \land F_2 \land \neg F_3$.
If the formula is satisfiable, the two queries are not equivalent, and the resulting model serves as a counterexample.
Even for this simple example, with a bound of 1 and only four SQL operations, the generated formula contains 179 AST nodes, including 78 variables and 101 operators.
In realistic scenarios, both the queries and the corresponding formulas grow substantially in complexity.

\section{Overview of \sys }
\label{sec:overview}

We develop a new complete verifier, \sys, that overcomes the limitations of existing list-based approaches.
Unlike prior methods, \sys employs a novel encoding based on U-expressions~\cite{chu2018udp}, which results in significantly simpler formulas.
Because no complete verification algorithm previously existed for U-expressions, \sys introduces a new decision procedure for reasoning about the generated formulas.
Beyond improved verification efficiency, \sys also supports a broader range of SQL features than existing verifiers.

\noindent
\textbf{U-expressions.}
\revised{Introduced by Chu et al.~\cite{chu2018udp}}, a U-expression is an algebraic representation of a query as a function $f(t)$ mapping each tuple $t$ to its multiplicity in the query result.
U-expressions capture a wide range of SQL features, including aggregate functions, \outerj, and \union~\cite{chu2018udp,sigmod2024sqlsolver,wang2024qed}.
They also support features beyond previous approaches, such as \intersect and \excepts.

%

%
%
%
%
%
%
%
%
%
%

%
\begin{table*}[t]
\caption{The terms/operators of U-expressions.}
\label{tab:uexpr-terms}
\centering
\begin{tabular}{|l|l|l|l|}
\hline
\multicolumn{1}{|c|}{\begin{tabular}[c]{@{}c@{}}\textbf{Term/}\textbf{Operator}\end{tabular}} 
& \multicolumn{1}{c|}{\textbf{Concept}} 
& \multicolumn{1}{c|}{\begin{tabular}[c]{@{}c@{}}\textbf{Example Query}\end{tabular}} 
& \multicolumn{1}{c|}{\begin{tabular}[c]{@{}c@{}}\textbf{Example} \textbf{U-expression} $f(t)$\end{tabular}} \\
\hline
$\llbracket \mathit{R} \rrbracket(t)$
& \begin{tabular}[c]{@{}l@{}} $\llbracket \mathit{R} \rrbracket(t)$ is a function that returns\\ the multiplicity of the tuple $t$\\ in the relation $R$.\end{tabular} 
& \texttt{{\color{awesome}select} $\ast$ {\color{awesome}from} R}
& $  \llbracket \mathit{R} \rrbracket(t)$ \\ \hline 
{[}b{]} 
& \begin{tabular}[c]{@{}l@{}} {[}b{]} returns 1 if the predicate $b$ is\\ true. Otherwise, it returns 0.
\end{tabular}
& \begin{tabular}[c]{@{}l@{}}  \texttt{{\color{awesome}select} $\ast$ {\color{awesome}from} R} \\ \texttt{{\color{awesome}where} R.x < 500} \end{tabular}
& $   \llbracket \mathit{R} \rrbracket(t) \times [t.x < 500]$ \\ \hline
$\sum_{t}e(t)$
& \begin{tabular}[c]{@{}l@{}} $\sum_{t}e(t)$ returns the sum of $e(t_i)$ for\\ each possible tuple $t_i$, where $e$ is a\\ U-expression in the type of $\mathit{tuple} \rightarrow \mathbb{N}$.\\ Since there may be unbounded $t_i$,\\ $\sum_{t}e(t)$ is called an unbounded summation. \end{tabular}
& \begin{tabular}[c]{@{}l@{}} \texttt{{\color{awesome}select} R.x {\color{awesome}from} R} \end{tabular} 
& \begin{tabular}[c]{@{}l@{}} $  \sum_{t_1}([t = t_1.x ]\times \llbracket \mathit{R} \rrbracket(t_1))$\end{tabular}
\\ \hline
\end{tabular}
\normalsize
\end{table*}
 
The main terms/operators of U-expressions are summarized in \Cref{tab:uexpr-terms}.
Each term returns a non-negative integer value and is connected by ``$\times$'' or ``$+$'' in the \uexp. 
Specifically, for each relation name $R$, there is a pre-defined function $\llbracket R \rrbracket(t)$ that returns the multiplicity of tuple $t$ in the relation $R$. 
For each predicate $b$, the \uexp $[b]$ is either 1 ($b$ is true) or 0 ($b$ is false).
For brevity, we omit $\llbracket \rrbracket$ in \uexps if there is no ambiguity. 
To illustrate U-expressions, we encode several core SQL features following the algorithm in \revised{\sqlsolver~\cite{sigmod2024sqlsolver}}.
\begin{itemize}
\item \sel: $Sel_p(Q)$ denotes a selection over the sub-query $Q$ with predicate $p$.  
Assume the U-expression of $Q$ is $f_l$.  
The U-expression for the selection is 
\[
f(t) \coloneqq f_l(t) \times [p(t)].
\]  
This preserves the multiplicity of $t$ if it satisfies $p$, and sets it to 0 otherwise.

\item \revised{\proj}: $Proj_a(Q)$ denotes returning the attributes $a$ from the result of the sub-query $Q$.  
Assume the U-expression of $Q$ is $f_l$.  
The U-expression is
\[
f(t) \coloneqq \sum_{t_l} f_l(t_l) \times [t = t_l.a].
\]  
The unbounded summation $\sum_{t_l}$ aggregates multiplicities of all tuples $t_l$ whose projected attributes $a$ match $t$.

\item \unionall: $UnionAll(Q_l, Q_r)$ denotes the union of the sub-queries $Q_l$ and $Q_r$.  
Assume the U-expressions of $Q_l$ and $Q_r$ are $f_l$ and $f_r$, respectively.  
The U-expression for the union is
\[
f(t) \coloneqq f_l(t) + f_r(t),
\]  
summing the multiplicities of $t$ from both sub-queries.
\end{itemize}

\noindent
\textbf{Basic idea of \sys.}
Our complete SQL equivalence verifier, \sys, is based on U-expressions.
Given two queries $Q_1$ and $Q_2$, \sys first translates them into U-expressions $f_1(t)$ and $f_2(t)$, respectively.
It then encodes the formula $\exists t. f_1(t) \neq f_2(t)$ into first-order logic, which can be automatically checked by an SMT solver.
If the formula is satisfiable, there exists a tuple whose multiplicities differ between the two query results; the SMT solver returns a model that can be used to construct a counterexample database.

\noindent
\textbf{Motivating Example.}
We now show how \sys disproves the queries in \Cref{fig:motivating-example} and compare the resulting formula with that of existing works.
First, the U-expressions for the queries $Q_1$ and $Q_2$ are $f_1(t)$ and $f_2(t)$, respectively.
\begin{equation*}
\begin{aligned}
f_1(t) &\coloneqq (hotel(t) \times [t.star < 4]) + (hotel(t) \times [t.price < 500])\\
f_2(t) &\coloneqq (hotel(t) \times [t.star < 4]) \\
\end{aligned}
\end{equation*}
Specifically, $f_1(t)$ is the addition of two sub-expressions, which represents the multiplicity of $t$ in each sub-query.
Then, \sys converts $\exists t. f_1(t) \neq f_2(t)$ to a FOL formula.
According to the schema of the {\tt{hotel}} table, $t$ is a tuple with three integer variables $(id_1, star_1, price_1)$.
The resulting FOL formula is
\begin{equation*}
\begin{aligned}
&\Big((hotel(id_1, star_1, price_1) \times int(star_1 < 4)) + \\
& (hotel(id_1, star_1, price_1) \times int(price_1 < 500))\Big)\\
\neq& (hotel(id_1, star_1, price_1) \times int(star_1 < 4))\\
\end{aligned}
\end{equation*}
Finally, an SMT solver checks the formula and reports that the queries are not equivalent.

In this example, \sys generates a formula with only 26 AST nodes, consisting of 15 variables and 11 operators, which is substantially smaller than the 179-node formula produced by a list-based verifier.
Our evaluation of test cases derived from Spider shows that 
\sys achieves a median verification time reduction of $83.1\%$ compared to \polygon.

\noindent
\textbf{Challenges.}
Directly applying U-expressions to construct a complete verifier is a non-trivial task.
First, U-expressions alone cannot capture queries with ordered bag semantics.
Second, encoding the unbounded summations in U-expressions into first-order logic is challenging, since FOL operators cannot express sums over unbounded terms.
We address these issues with an ordered-bag encoding (\Cref{sec:vc}) and a complete procedure for unbounded summations (\Cref{sec:approach}).
\section{Order Semantics Encoding}
\label{sec:vc}

We propose \emph{order semantics encoding} to fill the semantics gap between U-expressions and SQL queries that sort tuples. The order semantics of SQL queries can be involved by \limit, \off, and \order clauses. To handle order semantics, our approach consists of two parts: (1) SQL query preprocessing, which eliminates all \limit and \off clauses; and (2) query equivalence encoding, which handles the order semantics involved by \order.
The proof for the correctness of handling \limit/\off/\order can be found in Section 2 of supplementary material~\cite{supplementary}.

\subsection{SQL Query Preprocessing}
\label{subsec:preprocess}

The method of eliminating \limit and \off clauses is to transform a SQL query with these clauses into another new query without these clauses, while preserving the query semantics.
\limit and \off clauses select tuples whose rankings are within the specific range.
Without loss of generality, assume the tuples are sorted based on their values of attribute $a$ in descending order.
Our key insight is that the ranking of each tuple can be calculated by aggregating the multiplicity of other tuples that have a larger attribute $a$.
The calculation can be naturally implemented by the aggregate function ``sum''.
Thus, we can produce the result of \limit and \off clauses with aggregate functions.
Although we can completely remove \limit and \off clauses, it is noteworthy that we cannot completely remove \order clauses in general.
Otherwise, the query after transformation will have only bag semantics, which will be non-equivalent to the original query with order semantics.

\begin{figure}[t]
\begin{lstlisting}[language=SQL,breaklines=true,escapeinside={(*}{*)}]
-- Rule 1
(*$Q_1$*): R order by R.attr desc limit m offset n
(*$Q_2$*): select R1.* from
    (select R1.*, sum(case when R1.attr < R2.attr then 1 else 0 end) as s
     from R as R1 full join R as R2
     group by R1.attr
     having s < m+n and s >= n)
    order by R1.attr desc
\end{lstlisting}
\begin{lstlisting}[language=SQL,breaklines=true,escapeinside={(*}{*)}]
-- Rule 2
(*$Q_3$*): R order by R.attr asc limit m offset n
(*$Q_4$*): select R1.* from
    (select R1.*, sum(case when R1.attr > R2.attr then 1 else 0 end) as s
     from R as R1 full join R as R2
     group by R1.attr
     having s < m+n and s >= n)
    order by R1.attr asc
\end{lstlisting}
\caption{The preprocessing rules that remove \limit and \off in the given SQL query. Each rule transforms the first SQL query into the second SQL query; the two queries are equivalent.}
\label{fig:preprocess}
\end{figure}
 
\Cref{fig:preprocess} illustrates two preprocessing rules designed to remove all \limit and \off clauses from a query generally.
The first rule is designed to handle descending order, while the second rule is for ascending order.
For the first rule, the original query $Q_1$ first sorts the tuples in the result of $R$ by attribute \texttt{attr} in descending order, then skips the first $n$ tuples and returns the next $m$ tuples.
Note that $R$ can be a sub-query or a table.
The SQL query $Q_2$ achieves the same result by using a self-join and the aggregate function ``sum''.
Specifically, for each tuple in $R$, the aggregate function ``sum'' counts how many tuples in $R$ have larger values in attribute \texttt{attr}.
Based on this count, it filters out tuples to retain only those that would appear between positions $n+1$ and $m+n$ in the sorted sequence.
Then, the result is sorted by the attribute \texttt{attr}.
Note that the rule works when the attribute \texttt{attr} is unique.
Otherwise, $Q_1$ is ambiguous because the order of tuples with the same value in \texttt{attr} is not defined.
\sys does not consider ambiguous queries because their semantics are nondeterministic and difficult to formulate precisely.
\revised{
The second rule for ascending order is symmetric, which counts tuples with smaller rather than larger values.
NULL ordering can be captured similarly.
}

After removing all \limit and \off, \sys further removes \order in some subqueries (e.g., FROM subquery).
It does not change the query result, because SQL ignores such ORDER BY according to SQL standard and mainstream database manuals~\cite{snowflakeorderby,wang2022wtune}. In other cases, \sys promotes ORDER BY from sub-queries to the outermost level, which does not change the query semantics either.
Finally, each query will have at most one \order clause at the outermost level.

\subsection{Query Equivalence Encoding}
\label{subsec:encodeeq}

Given two preprocessed SQL queries $Q_1$ and $Q_2$, \sys further encodes their equivalence.
Specifically, \sys checks whether $Q_1$ and $Q_2$ have \order at the outermost level.
\sys uses different methods to encode SQL pairs into formulas for three possible cases: (1) both queries do not have \order; (2) only one query has \order; (3) both queries have \order.
If the resulting formula is satisfiable, the two SQL queries are not equivalent.
\revised{
Importantly, like existing works~\cite{he2024verieql,zhao2025polygon}, \sys does not depend on executing queries on actual data.
All entities in the formulas, including tuples and input tables, are symbolic, so the check considers every possible input database subject to the bound and integrity constraints.
If the formula is satisfiable, the SMT solver returns concrete table contents on which the queries produce different results; these contents are generated by the solver rather than supplied to \sys.
}

\vspace{3pt}
\noindent
\textbf{Case 1: Both queries do not have \order.}

In this case, two SQL queries are under bag semantics rather than order semantics.
Suppose $f_1$ and $f_2$ are the U-expressions modeling the two SQL queries, respectively.
The formula for verification is as follows:
\begin{equation*}
\label{eq:vc-no-order}
\begin{aligned}
& \exists t. f_1(t) \neq f_2(t) \\
\end{aligned}
\end{equation*}
The formula specifies that there is a tuple $t$ such that its multiplicities in the result of $Q_1$ and $Q_2$ are different.
If the formula is proved to be unsatisfiable, it means the SQL queries are equivalent.
Otherwise, it means that there exists a tuple $t$ such that the multiplicities of $t$ in the two query results are different.
The two queries are non-equivalent and \sys will return a counterexample.
\sys proves the formula using a new complete decision procedure, which will be discussed in \Cref{sec:approach}.

\vspace{3pt}
\noindent
\textbf{Case 2. Only one query has \order.}

\underline{\textit{Step 1:}}
\sys takes two SQL queries $Q_1$ and $Q_2$ as the input.
Suppose $Q_1$ has \order and $Q_2$ does not have \order.
$f_2$ is the U-expression of $Q_2$.
\sys reasons about the following formula to check whether there exists a symbolic input database, within the bound, on which $Q_2$ produces two distinct tuples.
\begin{equation}
\begin{aligned}
\exists t_1, t_2. ((t_1\neq t_2) \land (f_2(t_1) >0) \land (f_2(t_2)>0))
\end{aligned}
\label{eq:vc:case2:eq1}
\end{equation}
If the formula is satisfiable, the SMT solver returns concrete table contents on which $Q_2$ produces two distinct tuples $t_1$ and $t_2$.
If $Q_1$'s result also has the two different tuples $t_1$ and $t_2$, the tuples in $Q_1$'s result are ordered but the tuples in $Q_2$'s result are unordered.
Two queries are non-equivalent.
If $Q_1$'s result does not have $t_1$ or $t_2$, the two queries are still non-equivalent.

\vspace{2pt}
\underline{\textit{Step 2:}}
If \Cref{eq:vc:case2:eq1} is unsatisfiable, then on every input database within the bound, $Q_2$ returns either an empty result or tuples with the same value.
For example, if the schema declares a primary key and the query specifies its exact value, the query returns at most one tuple on every valid input database.
Thus, if the sub-query followed by \order in $Q_1$ is equivalent to $Q_2$, $Q_1$ and $Q_2$ are also equivalent.
This is because ordering an empty result or equal-valued tuples does not change the result.
Otherwise, $Q_1$ and $Q_2$ are non-equivalent.
Based on the above analysis, \sys extracts the sub-query followed by \order in $Q_1$.
Suppose the extracted sub-query is $Q_1'$ and the U-expression of $Q_1'$ is $f_1$.
\sys checks whether the following formula is satisfiable.
\begin{equation*}
\begin{aligned}
\exists t. f_1(t)\neq f_2(t)
\end{aligned}
\end{equation*}
The formula specifies that there is a tuple $t$ such that its multiplicities in the result of $Q_1'$ and $Q_2$ are different.
If it is satisfiable, \sys will return the counterexample.
Otherwise, two queries are equivalent.

\vspace{3pt}
\noindent
\textbf{Case 3. Both queries have \order.}

\underline{\textit{Step 1:}}
\sys first checks whether the SQL queries $Q_1$ and $Q_2$ have equivalent sub-queries.
If not, $Q_1$ and $Q_2$ are non-equivalent.
Suppose $Q_1'$ and $Q_2'$ are the sub-queries of $Q_1$ and $Q_2$, respectively. $f_1$ and $f_2$ are the corresponding U-expressions of $Q_1'$ and $Q_2'$.
If $\exists t. f_1(t) \neq f_2(t)$ is satisfiable, there exists a tuple $t$ such that $Q_1'$ and $Q_2'$ are non-equivalent.
\sys will return the counterexample for $Q_1'$ and $Q_2'$, which is also the counterexample for the original queries $Q_1$ and $Q_2$.
Because \order only sorts tuples without changing tuples.

\vspace{2pt}
\underline{\textit{Step 2:}}
If the sub-queries of $Q_1'$ and $Q_2'$ are equivalent, \sys will further check whether $Q_1$ and $Q_2$ sort the tuples in the same order.
In other words, \sys needs to check whether there exist two tuples $t_1$ and $t_2$ such that: 1) $t_1$ is before $t_2$ in the result of $Q_1$, and 2) $t_1$ is after $t_2$ in the result of $Q_2$.
Assume that $Q_1$ and $Q_2$ sort the tuples produced by their sub-queries according to the attribute $a_1$ with order $dir_1$ and $a_2$ with order $dir_2$, respectively. 
Here $dir_1$ and $dir_2$ can be either ascending order or descending order.

\begin{itemize}
    \item If $dir_1$ is the same as $dir_2$, the formula for verification is
    \begin{equation*}
    \begin{aligned}
\exists t_1, t_2. ((f_1(t_1) > 0) \land (f_1(t_2) > 0) \land (t_1.a_1 > t_2.a_1) \land (t_1.a_2 < t_2.a_2))
    \end{aligned}
    \end{equation*}
It means that there exist two tuples $t_1$ and $t_2$ such that: 1) $t_1$ is in the result of two SQL queries ($f_1(t_1) > 0$); 2) $t_2$ is in the result of two SQL queries ($f_1(t_2) > 0$); 3) their relative orders in the results of $Q_1$ and $Q_2$ are different ($(t_1.a_1 > t_2.a_1) \land (t_1.a_2 < t_2.a_2)$).
If the above formula is satisfiable, $Q_1$ and $Q_2$ are non-equivalent.
Otherwise, they are equivalent.
Note that as we have proved the equivalence between sub-queries of $Q_1$ and $Q_2$ (\emph{i.e.}, U-expressions $f_1$ and $f_2$ are equivalent),
$(f_1(t_1) > 0) \land (f_1(t_2) > 0)$ alone can infer that both $t_1$ and $t_2$ exist in the results of both $Q_1$ and $Q_2$.

\item If $dir_1$ is different from $dir_2$, the formula for verification is
\begin{equation*}
\begin{aligned}
\exists t_1, t_2. ((f_1(t_1) > 0) \land (f_1(t_2) > 0) \land (t_1.a_1 > t_2.a_1) \land (t_1.a_2 > t_2.a_2))
\end{aligned}
\end{equation*}
The meaning of the above formula is similar to the previous formula except for $t_1.a_2 > t_2.a_2$.
This is because one query now sorts tuples in ascending order and the other sorts them in descending order.

\end{itemize}
\section{Unbounded Summation Expansion}
\label{sec:approach}

This section introduces how we enable the conversion of U-expressions with unbounded summations to pure FOL formulas without losing completeness.
Our basic idea is to integrate bounded verification~\cite{Kokologiannakis2021genmc,Kokologiannakis2019lock,schuele2007bmc,bieretacas1999,Oberhauser2019vsync} into U-expressions. We propose some rules to expand unbounded summations into bounded summations with specific symbolic tuples. 

\subsection{Basic Idea}
\label{subsec:basid-idea}

The challenge of handling unbounded summations stems from the unknown number of distinct tuples in each table.
An unbounded summation needs to enumerate all possible tuples because it is unknown which tuples can make the summation body not zero.
However, we observe that if the number of distinct tuples in each table is known, the number of tuples that make the summation body non-zero will be bounded, and thus the unbounded summation can be expanded to a bounded summation.
\begin{algorithm}[t]
\setcounter{AlgoLine}{0}
\caption{Decision Procedure}
\label{alg:dp}
\footnotesize
    \let\oldgets\gets
    \renewcommand{\gets}{\textbf{:= }}
    \tt
    \textbf{Input:} A formula $F$ with first-order logic terms and U-expressions.\\
    \textbf{Output:} The satisfiability of the formula.\\
    \underline{Prove}($F$): \\
    \ \ $b$ \gets GetBoundConfig() \label{alg:dp:readbound}\\
    \ \ $M$ \gets ConstructTableToTuples($F$, $b$) \label{lineno:dp:tabletotuple} \\
    \ \ $F_1$ \gets ExpandUnboundedSummations($F$, $M$) \label{alg:dp:expandsum} \\
    \ \ $result$ \gets SMTSolve($F_1$) \\
    \ \ \textbf{if} $result$ = UNSAT \textbf{then} \\
    \ \ \ \ \textbf{return} EQ \\
    \ \ \textbf{else} \\
    \ \ \ \ \textbf{return} $result$ \\
    \let\gets\oldgets
\normalsize
\end{algorithm}
 
Based on this insight, we propose \emph{unbounded summation expansion}, which combines U-expressions with bounded verification to better model the summation semantics of U-expressions and facilitates the capability of complete verification for SQL equivalence.
\Cref{alg:dp} shows the overall workflow of our approach.
The basic idea is to set the maximum number of distinct tuples in each table as a bound and perform verification under the bound.
The value of the bound can be read from a configuration file by invoking $\textit{GetBoundConfig}$ (Line~\ref{alg:dp:readbound}).
The workflow consists of the following steps.
First, we construct a mapping $M$ based on the bound $b$, which maps a table name to a list of $b$ distinct symbolic tuples.
Second, based on $M$, we replace all unbounded summations in the formula $F$ with bounded summations (detailed in \Cref{subsec:summation-expansion}).
Finally, we use an off-the-shelf SMT solver to check the satisfiability of the resulting pure first-order logic formula.
If the verification condition is satisfiable under a certain bound, the two SQL queries are not equivalent.
SMT solvers will provide a counterexample.
Otherwise, users can iteratively increase the bound and rerun the verifier until it finds a counterexample or the verification times out.
To reason about the equivalence of U-expressions under integrity constraints, \sys adopts a method similar to those in prior work~\cite{he2024verieql}.
Assume the encoding of query equivalence without integrity constraints is $F$.
It first encodes the integrity constraints into a first-order logic formula $P$ without quantifiers.
Then, \sys checks the satisfiability of $P \land F$ with \Cref{alg:dp}.
Since the encoding of integrity constraints is similar to that in prior work~\cite{he2024verieql,zhao2025polygon}, we omit the details here.

Note that although bounded verification has been adopted in prior works on SQL query verification~\cite{he2024verieql,zhao2025polygon}, their bound has a different meaning from ours.
In prior works, the bound means the size of the entire table, while our bound is set on the number of distinct tuples rather than all tuples.
For example, if the bound is set to one, existing works assume that each table contains at most one tuple, while we assume that each table contains at most one distinct tuple but can have multiple tuples with the same value.
Thus, with the same bound, \sys may find counterexamples that existing works cannot find.
For example, consider the SQL queries ``\sql{select} a \sql{from} r'' and ``\sql{select distinct} a \sql{from} r''.
When the bound is one, existing verifiers cannot find counterexamples because they assume the table r has at most one tuple. Both queries will return the tuple.
However, with the same bound, \sys can find the counterexample, which is a table $r$ consisting of two tuples with the same value.
Then, ``\sql{select} a \sql{from} r'' will return two tuples, while ``\sql{select distinct} a \sql{from} r'' will return only one because ``\sql{distinct}'' removes duplicate tuples.
Thus, \sys can prove that these queries are non-equivalent.

\begin{figure}[t]

\begin{lstlisting}[language=SQL,breaklines=true,escapeinside={(*}{*)}]
create table stu(id int, age int);
(*$Q_1$*): select id from stu where age > 18
(*$Q_2$*): select id from stu where age + 4 > 22
\end{lstlisting}
\normalsize
\leftline{Their corresponding \uexps are:}

\begin{equation*}
\begin{aligned}
f_1(t) \coloneqq 
&\sum_{t_1}([t=t_1.id] \times stu(t_1) \times [t_1.age > 18]) \nonumber \\
f_2(t) \coloneqq 
&\sum_{t_2}([t=t_2.id] \times stu(t_2) \times [t_2.age + 4 > 22]) \nonumber
\end{aligned}
\end{equation*}
\caption{An example of two equivalent SQL queries along with their U-expressions.}
\label{fig:sum-example}
\end{figure}
 
By determining the number of distinct tuples in each table, our approach can accurately (\emph{i.e.}, without approximation~\cite{levatich2020solving}) represent the unbounded summations with the terms that can be formulated into first-order logic. This means that the approach preserves verification completeness.
Our approach iteratively repeats the following three steps until all unbounded summations have been expanded into bounded summations.
First, the algorithm finds the innermost unbounded summation.
A summation $\sum_{t} e(t)$ is innermost if $e(t)$ does not have any summation.
Second, the algorithm will construct a bounded summation domain by recursively analyzing the summation body.
Finally, assume the summation domain for $\sum_{t} e(t)$ is $\{ t_1, \ldots, t_n \}$.
The summation $\sum_{t} e(t)$ in the original U-expression will be expanded into $e(t_1) + \ldots + e(t_n)$.

For each unbounded summation $\sum_{t} e(t)$, it is intuitive to set the summation domain as a bounded set consisting of new symbolic tuples.
However, this intuitive method has two problems.
First, it is hard to determine the size of the summation domain.
The domain size should be larger than the number of distinct tuples that can make the summation body non-zero.
Otherwise, the original summation will not be equivalent to the bounded summation.
However, an overly large domain may contain unnecessary tuples, resulting in overly long formulas and slow verification.
Second, the above method needs to introduce a constraint with quantifiers, which specifies that any tuple outside the domain will make the summation body zero.
Specifically, for the unbounded summation $\sum_{t} e(t)$ and summation domain $\{t_1, \ldots, t_n\}$, the constraint is $\forall t. ((t \neq t_1 \land \ldots \land t \neq t_n) \rightarrow (e(t) = 0))$.
Quantifiers are notoriously difficult for SMT solvers to handle.
However, omitting the quantifiers loses completeness, which means that an unsatisfiable formula may be incorrectly judged as satisfiable.
For example, \Cref{fig:sum-example} shows two equivalent queries and their U-expressions $f_1$ and $f_2$.
To prove their equivalence, we need to prove that $\exists t. f_1(t) \neq f_2(t)$ is unsatisfiable.
Assume that the summation domain of $\sum_{t_1}$ in $f_1$ is $\{a_1, a_2, \ldots, a_m\}$ and that the summation domain of $\sum_{t_2}$ in $f_1$ is $\{b_1, b_2, \ldots, b_n\}$.
Then, the formula will be as follows.
\begin{equation*}
\begin{aligned}
\exists t. &\sum_{i=1}^{m}([t=a_i.id] \times stu(a_i) \times [a_i.age > 18]) \\
&\quad \neq \sum_{i=1}^{n}([t=b_i.id] \times stu(b_i) \times [b_i.age + 4 > 22])
\end{aligned}
\end{equation*}
The original formula is unsatisfiable, while the above formula is satisfiable because the latter holds when each $students(a_i)$ is zero and each $students(b_i)$ is one.

\subsection{Bounded Summation Domain}
\label{subsec:summation-expansion}

To address the challenge, the basic idea is to precisely find each tuple that can make the summation body non-zero rather than directly using new symbolic tuples.
Before going into the details, we first consider the example in \Cref{fig:sum-example}.
Assume that the bound is 1, which means the table has at most one distinct tuple.
We use $x$ to represent the tuple.
Then, it is clear that the summation domain of $\sum_{t_1}$ in $f_1$ is $\{x\}$.
Because $stu(t_1)$ can be non-zero only when $t_1$ equals $x$, which infers that the summation body can be non-zero only when $t_1$ equals $x$.
The summation domain of $\sum_{t_2}$ in $f_2$ is also $\{x\}$ due to the same reason.
Then, $\exists t. f_1(t) \neq f_2(t)$ will become the following formula.
\begin{equation*}
\begin{aligned}
\exists t. &([t=x.id] \times stu(x) \times [x.age > 18])\\
&\quad \neq ([t=x.id] \times stu(x) \times [x.age + 4 > 22])
\end{aligned}
\end{equation*}
The formula is unsatisfiable because $[x.age > 18]$ is equivalent to $[x.age + 4 > 22]$, which infers that the two U-expressions are equivalent.

\begin{table}[t]
\caption{The rules of constructing a bounded summation domain $dom(e, t, M)$, which is a finite set covering all values of the tuple $t$ that can make $e$ non-zero. $M$ is a mapping from a table name to a set of symbolic tuples in the table, which is created at Line~\ref{lineno:dp:tabletotuple} in \Cref{alg:dp}.}
\centering
\begin{tabular}{|>{\centering\arraybackslash}p{0.57\linewidth}|>{\centering\arraybackslash}p{0.32\linewidth}|}
\hline
$e$  & $dom(e, t, M)$ \\
\hline
$r(t)$ & $M(r)$ \\ \hline
$[t = t']$ & $\{t'\}$ \\ \hline
$[IsNull(t)]$ & $\{NULL\}$ \\ \hline
$\|e_1\|$ & $dom(e_1, t, M)$ \\ \hline
$e_1\times \ldots\times e_i\times\ldots \times e_n$, where $dom(e_i, t, M) \neq \emptyset$ & $dom(e_i, t, M)$ \\ \hline
$[t.a_1 = v_1] \times \ldots \times [t.a_n = v_n] \times e'$, where $t = (a_1, \ldots, a_n)$ and $notuse(v_1, t)\land \ldots \land notuse(v_n, t)$ & $\{(v_1, \ldots, v_n)\}$ \\ \hline
$e_1 + e_2$, where $dom(e_1, t, M) \neq \emptyset$ and $dom(e_2, t, M)\neq \emptyset$ & $dom(e_1, t, M) \cup dom(e_2, t, M)$ \\ \hline
$\sum_{t_1} e_1(t_1)$, where $dom(e_1,t_1,M)=S$ and $S \neq \emptyset$ & $dom(\sum_{v\in S}e_1(v), t, M)$ \\ \hline
Others & $\emptyset$ \\
\hline
\end{tabular}
\label{fig:domain-rules}
\end{table}

Based on the above observation, we present rules to systematically construct \revised{a bounded summation domain}, as shown in \Cref{fig:domain-rules}.
Given unbounded summation $\sum_{t} e(t)$, our rules construct a bounded set of tuples denoted as $dom(e, t, M)$.
$M$ is a mapping from a table name to a set of symbolic tuples, which is created at Line~\ref{lineno:dp:tabletotuple} in \Cref{alg:dp}.
Each table $r$ only contains tuples in $M(r)$.
The purpose of $dom(e,t,M)$ is to return a finite set covering all values of $t$ such that $e(t)\neq0$.
It can contain $t$ s.t. $e(t)=0$ because this does not break $\sum_{t}e(t)=\sum_{v{\in}dom(e,t,M)}e(v)$.
If $e(t)$ is always zero or there are infinitely many $t$ such that $e(t)\neq0$, $dom(e,t,M)$ returns $\emptyset$.
These rules consider all possible structures of U-expressions and do not create any new symbolic variables.
Here is the detailed explanation:

\begin{itemize}

\item \emph{DOM-Rel:}
Assume $e$ is in the form of $r(t)$, where $r$ is an uninterpreted function that returns the multiplicity of $t$ in the table $r$.
$r(t)$ is non-zero if and only if $t$ is a tuple in the table $r$.
Thus, $dom(e, t, M)$ is $M(r)$.

\item \emph{DOM-Eq:}
Assume $e$ is in the form of $[t = t']$, where $t'$ is another symbolic tuple.
It is non-zero if and only if the predicate $t = t'$ is true.
Thus, $dom(e, t, M)$ is $\{t'\}$.

\item \emph{DOM-Null:}
Assume $e$ is in the form of $[IsNull(t)]$, where the predicate $IsNull$ means $t$ is NULL.
It is non-zero if and only if $IsNull(t)$ holds.
Thus, $dom(e, t, M)$ is $\{NULL\}$.

\item \emph{DOM-Squash:}
Assume $e$ is in the form of $\|e_1\|$, where $e_1$ is another U-expression.
It is non-zero if and only if $e_1$ is non-zero.
Thus, $dom(e, t, M)$ is equivalent to $dom(e_1, t, M)$.
\revised{
For example, $dom(\|stu(t)\|, t, M)$ is equivalent to $dom(stu(t), t, M)$, which is $M(stu)$.
This means that only tuples in the table $stu$ can make $\|stu(t)\|$ non-zero.
}

\item \emph{DOM-Mul1:}
Assume $e$ is the multiplication of U-expressions $e_1, \ldots, e_n$.
$e$ is not zero if and only if each $e_i \in \{ e_1, \ldots, e_n \}$ is not zero.
Thus, if $\exists{e_i}.(e=e_1\times\dots\times{e_i}\times\dots\times{e_n})\land(dom(e_i,t,M)\neq\emptyset)$, then $dom(e,t,M)=dom(e_i,t,M)$.
\sys checks each $e_i$ sequentially and picks the first $e_i$ s.t. $dom(e_i,t,M)\neq\emptyset$.
Note that although $dom(e, t, M)$ may contain some tuples that can make some other $e_j (i \neq j)$ zero, it does not corrupt completeness, because the rule will not miss any tuples that can make $e$ non-zero.
\revised{
For example, $dom(stu(t_1)\times[t_1.age>18],t_1,M)$ is equivalent to $dom(stu(t_1),t_1,M)$, which is $M(stu)$.
This means that only tuples in the table $stu$ can make $stu(t_1)\times[t_1.age>18]$ non-zero.
}

\item \emph{DOM-Mul2:}
Assume the tuple $t$ is a list of symbolic attributes $a_1, \ldots, a_m$ and $e$ is in the form of $[t.a_1 = v_1] \times \ldots \times [t.a_n = v_n] \times e'$.
Specifically, $t.a_i$ is the attribute $a_i$ of the tuple $t$.
$e', v_1, \ldots, v_n$ are U-expressions.
The rule returns ${(v_1, \dots, v_n)}$, because only the tuple $(v_1, \dots, v_n)$ can enable $e\neq0$.
Note that since the U-expression $e$ should not contain $t$ after expansion, the constructed summation domain also should not contain any attributes in $t$.
Thus, this rule also requires that each $v_i (1\le i \le n)$ must not use the attributes in $t$, which is represented by $notuse(v_1, t)\land \ldots \land notuse(v_n, t)$.
\revised{
For example, $dom([t_1.id=3]\times[t_1.age=18],t_1,M)$ is equivalent to $\{(3, 18)\}$.
This means that only the tuple $(3, 18)$ can make $[t_1.id=3]\times[t_1.age=18]$ non-zero.
}

\item \emph{DOM-Add}: 
Assume $e$ is in the form of $e_1 + e_2$, where $e_1$ and $e_2$ are two other U-expressions.
It is non-zero if and only if $e_1$ is non-zero or $e_2$ is non-zero.
Thus, $dom(e, t, M)$ is equivalent to $dom(e_1, t, M)\cup dom(e_2, t, M)$.
\revised{
For example, $dom([t_1=t] + [IsNull(t_1)],t_1,M)$ is equivalent to $\{t, NULL\}$, which means both $t$ and $NULL$ can make $[t_1=t] + [IsNull(t_1)]$ non-zero.
}
Note that it requires $(dom(e_1,t,M)\neq\emptyset) \land (dom(e_2, t, M)\neq\emptyset)$ rather than $(dom(e_1,t,M)\neq\emptyset) \lor (dom(e_2, t, M)\neq\emptyset)$.
For example, $dom(1+[t=0],t,M)$ should be $\emptyset$, meaning $1+[t=0]$ is always non-zero regardless of $t$. If the rule uses $\lor$ rather than $\land$, it would incorrectly return ${0}$, which further makes $dom((1+[t=0])\times{r(t)},t,M)$ equal ${0}$ rather than $M(r)$.

\item \emph{DOM-Sum}:
Assume $e$ is in the form of an unbounded summation $\sum_{t_1}e_1(t_1)$ and $S=dom(e_1,t_1,M)$. Then, $dom(e,t,M)=dom(\sum_{v{\in}S}e(v),t,M)$. It means \sys recursively handles the innermost unbounded summation.
\revised{
For example, assume $M(stu)=\{x_1\}$ and consider $dom(\sum_{t_1}(stu(t_1)\times[t=t_1.id]),t,M)$.
For the summation body $stu(t_1)\times[t=t_1.id]$, \emph{DOM-Rel} infers that $dom(stu(t_1),t_1,M)=M(stu)$, which is $\{x_1\}$.
Then, \emph{DOM-Mul1} infers that $dom(stu(t_1)\times[t=t_1.id],t_1,M)$ is also $\{x_1\}$.
Thus, \emph{DOM-Sum} rewrites the unbounded summation as $stu(x_1)\times[t=x_1.id]$, after which the remaining rules infer that $dom(stu(x_1)\times[t=x_1.id],t,M)$ is $\{x_1.id\}$.
This means that $\sum_{t_1}(stu(t_1)\times[t=t_1.id])$ is non-zero only when $t$ equals $x_1.id$.
}

\item \emph{DOM-Other}:
If $e$ does not match any of the above forms, such as constants or negations, $dom(e, t, M)$ is $\emptyset$.
\revised{
For example, $dom(1,t,M)=\emptyset$ because the constant $1$ is non-zero for every possible value of $t$.
}
\end{itemize}

To further illustrate the rules, we provide two more complex examples below.

\noindent
\textbf{Example 1.} Consider the unbounded summation $\sum_{t_1}([t=t_1.id] \times stu(t_1) \times [t_1.age > 18])$ of the U-expression $f_1(t)$ in \Cref{fig:sum-example}.
Assume the bound is 2 and $M(stu)$ is $\{x_1, x_2\}$, where $x_1$ and $x_2$ are two distinct symbolic tuples.
According to \emph{DOM-Rel}, $dom(stu(t_1), t_1, M)$ is $M(stu)$.
By further applying \emph{DOM-Mul1}, $dom([t=t_1.id] \times stu(t_1) \times [t_1.age > 18], t_1, M)$ is also $M(stu)$.
Then, $f_1(t)$ is transformed to $\sum_{t_1 \in \{x_1, x_2\}}([t=t_1.id] \times stu(t_1) \times [t_1.age > 18])$, which is a bounded summation.
Similarly, $f_2(t)$ is transformed to $\sum_{t_2\in \{x_1, x_2\}}([t=t_2.id] \times stu(t_2) \times [t_2.age + 4 > 22])$.
Finally, \sys will check the satisfiability of the following formula:
\begin{equation*}
\begin{aligned}
&\sum_{t_1 \in \{x_1, x_2\}}([t=t_1.id] \times stu(t_1) \times [t_1.age > 18]) \\
\neq& \sum_{t_2\in \{x_1, x_2\}}([t=t_2.id] \times stu(t_2) \times [t_2.age + 4 > 22])
\end{aligned}
\end{equation*}
We additionally add $x_1\neq x_2$ to specify that they are distinct tuples in the set $M(stu)$.
This formula is apparently unsatisfiable, which infers that two queries are equivalent. 

\noindent
\textbf{Example 2.} Consider the query ``\texttt{\sql{select} id \sql{from} (\sql{select} $\ast$ \sql{from} stu \sql{union all} \sql{select} $\ast$ \sql{from} stu)}'', which uses the same schema as in \Cref{fig:sum-example}.
This query is simple for illustration purpose.
The U-expression is \revised{$f(t) \coloneqq\sum_{t_1}((stu(t_1) + stu(t_1))\times [t=t_1.id])$}.
Assume the bound is 2 and $M(stu)$ is $\{x_1, x_2\}$.
By applying \emph{DOM-Rel}, $dom(stu(t_1), t_1, M)$ is $M(stu)$.
Then, \emph{DOM-Add} infers that $dom(stu(t_1) + stu(t_1), t_1, M)$ is $M(stu)\cup M(stu)$, which is $M(stu)$.
Since $dom(stu(t_1) + stu(t_1), t_1, M)$ is not $\emptyset$, \emph{DOM-Mul1} further infers that \revised{$dom((stu(t_1) + stu(t_1))\times [t=t_1.id], t_1, M)$} is $dom(stu(t_1) + stu(t_1), t_1, M)$, which is also $M(stu)$.
Finally, the unbounded summation becomes a bounded summation \revised{$\sum_{t_1\in \{x_1, x_2\}}((stu(t_1) + stu(t_1))\times [t=t_1.id])$}.

\subsection{Theoretical Analysis}
\label{subsec:theorem-proving}

For any unbounded summation in the U-expression of a given SQL query, the rules in \Cref{fig:domain-rules} can always infer a non-empty and bounded set, which can help transform it to an equivalent bounded summation.
Note that although it is easy to construct a summation $\sum_{t} e(t)$ such that $dom(e, t, M)$ is $\emptyset$, such summation will not exist in U-expressions of valid SQL queries.
Here are theorems that formalize the above statements.
\revised{
Due to space limitation, we omit the proof, which can be found in Section 3 of supplementary materials~\cite{supplementary}.
}

\begin{theorem}
For any unbounded summation $\sum_{t} e(t)$ in the U-expression of a SQL query and a mapping $M$, if the rules in \Cref{fig:domain-rules} generate a set $S \coloneqq dom(e, t, M)$, $S$ is bounded.
\label{theo:bound-sum}
\end{theorem}

\begin{theorem}
For any U-expression of a SQL query $Q$, a bound $b$, and a mapping $M$, the following properties hold:

(1) For any unbounded summation $\sum_{t}e(t)$ in the U-expression, the rules in \Cref{fig:domain-rules} will generate a set $S\coloneqq{dom}(e,t,M)$ such that $S\neq\emptyset$ and $\sum_{t}e(t)=\sum_{v\in{S}}e(v)$.

(2) If replacing all unbounded summations with bounded summations (i.e., replacing $\sum_{t}e(t)$ with $\sum_{v\in{S}}e(v)$ above) via the rules in \Cref{fig:domain-rules} will generate a new U-expression $f'(t)$, then $dom(f',t,M)\neq\emptyset$ and $\forall{v}.((f'(v)\neq0)\rightarrow(v{\in}dom(f',t,M)))$.
\label{theo:domain-induction}
\end{theorem}

\begin{theorem}
For any U-expression $f(t)$ of a SQL query, a bound $b$, and a mapping $M$, if $f(t)$ becomes $f'(t)$ after replacing unbounded summations with bounded summations based on rules, then $f(t)=f'(t)$.
\label{theorem:sum}
\end{theorem}
\section{Soundness and Completeness}
\label{sec:theory}

This section provides some theorems about the soundness and completeness of our verification method.
The proof can be found in Section 4 of supplementary materials~\cite{supplementary}.

\begin{theorem}[Guarantee of Completeness]
For any queries $Q_1$, $Q_2$, integrity constraints $\textit{IC}$, and a bound $b$, if \sys finds a counterexample, then $Q_1 \neq Q_2$.
\label{theorem:5-1}
\end{theorem}

\begin{theorem}
There exist non-equivalent queries $Q_1$, $Q_2$, integrity constraints $\textit{IC}$, and a bound $b$ such that \sys cannot find counterexamples.
\label{theorem:5-2}
\end{theorem}

\begin{theorem}
For any queries $Q_1$, $Q_2$, integrity constraints $\textit{IC}$, and a bound $b$, if there exists database data $\textit{DB}$ such that 1) each table in $\textit{DB}$ has at most $b$ distinct rows, and 2) $Q_1(DB) \neq Q_2(DB)$, then \sys will return a counterexample. We assume the SQL queries do not have SQL features unsupported by \sys, and SMT solvers can solve the generated SMT formula.
\label{theorem:5-3}
\end{theorem}

\begin{theorem}
(1) For any queries $Q_1$ and $Q_2$, integrity constraint $\textit{IC}$, and a bound $b$, if \polygon finds a counterexample, \sys can find a counterexample. We assume that SMT solvers can successfully solve the generated formulas.

(2) There exist queries $Q_1$ and $Q_2$, integrity constraint $\textit{IC}$, and a bound $b$ such that \sys can find a counterexample and \polygon can't.
\end{theorem}
\section{Implementation}
\label{sec:impl}

\sys reuses SQL parser and U-expression encoder of \sqlsolver~\cite{sigmod2024sqlsolver}.
However, \sys adds ordered bag semantics and unbounded-summation expansion modules.
Finally, \sys encodes bounded U-expressions as a pure FOL formula and checks it with Z3~\cite{z3source}.
A satisfiable formula yields a counterexample distinguishing the queries; otherwise, they are equivalent.
\section{Evaluation}
\label{sec:eval}

In the evaluation, we answer the following questions: 1) Does \sys have better verification capability than prior work? 2) Does \sys have faster verification speed than prior work? 3) What are the main reasons for \sys's advantages?

\subsection{Experimental Setup}
\label{subsec:setup}

\noindent
\textbf{Baseline.}
To answer the above questions, we compare \sys with \verieql~\cite{he2024verieql} and \polygon~\cite{zhao2025polygon}, which are the state-of-the-art complete verifiers for SQL equivalence.
\revised{
The evaluation does not compare \sys with sound verifiers because they are incomplete and therefore are not helpful baselines for counterexample generation.
For example, the recent sound verifier \sqlsolver cannot generate counterexamples for any query pair, regardless of whether the queries are equivalent.
}
\revised{
For an apples-to-apples comparison, we do not compare \sys with testing tools for SQL equivalence.
Note that prior evaluations have compared our baselines, \verieql and \polygon, with various testing tools~\cite{coelho2013datafiller,chandra2019icde,chandra2015data,jeroen2018icde}.
The results demonstrate that these verifiers have a stronger ability to identify non-equivalent query pairs than testing tools.
Thus, we do not repeat the comparison with testing tools and instead compare \sys directly with \verieql and \polygon.
Moreover, testing tools do not provide the theoretical guarantees offered by complete verifiers.
}

\noindent
\textbf{Benchmark.}
\revised{To evaluate the verification capability and efficiency of each approach, we use three benchmarks.}
The first benchmark contains 997 pairs of SQL queries derived from Spider~\cite{yu2018spider}, which is a popular benchmark for the text-to-SQL task.
Each test case in Spider requires an LLM to translate a natural-language question about a database into a SQL query.
Since Spider is widely used to evaluate LLM-based text-to-SQL models, it represents realistic SQL queries and covers commonly used SQL features.
We use the translation results of DAIL-SQL~\cite{gao2024dailsql}, the top model on the Spider leaderboard~\cite{spider}.
Then, we pair the generated query with the corresponding correct answer provided by Spider to form a test case of equivalence verification.
The benchmark contains both equivalent and non-equivalent SQL query pairs.
To check whether two queries in each test case are indeed equivalent or non-equivalent, we combine sound verifiers (\emph{e.g.}, \sqlsolver~\cite{sigmod2024sqlsolver}) and complete verifiers (\emph{e.g.}, \polygon).
If a pair of SQL queries is verified to be non-equivalent by complete verifiers or equivalent by sound verifiers, we directly adopt the verification result.
Otherwise, we manually check the queries.
Among the original 1,034 query pairs, 726 are equivalent and 271 are non-equivalent.
We discard the other 37 pairs because they are ambiguous for the reason described in \Cref{subsec:preprocess}.

\revised{
The second benchmark contains 127 query pairs derived from the \spark benchmark used by \sqlsolver~\cite{sigmod2024sqlsolver}.
Seventeen of the 127 cases use SQL features unsupported by \sys, such as window functions.
The original benchmark contains equivalent pairs extracted from Spark SQL optimizer rewrites.
To evaluate counterexample generation, we use an LLM to mutate one query in each pair, making the pair non-equivalent.
We confirm non-equivalence by finding a counterexample for every pair.
The number of cases containing SQL features unsupported by \sys increases from 14 to 17 after mutation.
This increase indicates that the mutation does not construct only favorable cases and that the benchmark reflects the limitations of \sys.
}

To further evaluate the verification efficiency on more complex verification tasks, we adopt another benchmark with 1,000 test cases derived from \cubes, which was crafted by \polygon~\cite{zhao2025polygon}.
We do not consider the benchmarks used in \verieql, such as Calcite, because most cases in the Calcite benchmark are equivalent~\cite{he2024verieql}.
\verieql finds only 4 pairs of non-equivalent queries from this benchmark.
Thus, Calcite is not suitable to reflect the benefit of finding counterexamples faster.
Unlike the Spider and \spark benchmarks, each test case in \cubes contains 100 SQL queries rather than two SQL queries.
For each test case, the verifier needs to encode the semantics of 100 SQL queries into only one formula and find a counterexample that can classify these queries into different equivalence classes.
It means that when running the 100 SQL queries on the counterexample, some queries will produce the same result, while the other SQL queries will produce a different result.
This benchmark demonstrates that, in addition to the traditional task of verifying equivalence between two SQL queries, \sys can reason about multiple SQL queries.
\polygon also crafts another benchmark, each test case of which contains 50 SQL queries.
However, we do not use it in the evaluation because it is similar to the above benchmark and each test case contains fewer SQL queries.
Note that the \cubes benchmark was crafted by \polygon and uses only SQL features supported by \polygon and \sys.
Thus, it is mainly used to evaluate the verification efficiency.

\noindent
\textbf{Testbed.}
We run all experiments on a machine with 16 vCPUs and 32 GiB of memory.
It uses a third-generation Intel Xeon Scalable processor and runs 64-bit Ubuntu 24.04 Server.
Both \sys and \polygon use Z3 4.15.4~\cite{z3source,de2008z3} as the underlying SMT solver.
\sys reuses the SQL parser of Apache Calcite~\cite{begoli2018apache,ApacheCalcite} to parse SQL queries.

\subsection{Verification Capability}
\label{subsec:eval:capability}

\begin{table}[t]
\caption{The comparison among \sys, \verieql, and \polygon on the Spider and \spark benchmarks. The table shows the total number of test cases proved to be non-equivalent under each bound.}
\centering
\begin{tabular}{@{}llrrrr@{}}
\toprule
\multirow{2}{*}{\textbf{Benchmark}} & \multirow{2}{*}{\textbf{Verifier}} & \multicolumn{4}{c}{\textbf{Bound}} \\
\cmidrule(l){3-6}
& & \textbf{1} & \textbf{2} & \textbf{3} & \textbf{4} \\
\midrule
Spider & \sys     & 191 & 257 & 265 & 269 \\
       & \verieql & 70  & 79  & 81  & 81  \\
       & \polygon & 105 & 138 & 140 & 140 \\
\midrule
\revised{\spark} & \revised{\sys}     & \revised{63} & \revised{90} & \revised{110} & \revised{110} \\
                  & \revised{\verieql} & \revised{41} & \revised{54} & \revised{55}  & \revised{55}  \\
                  & \revised{\polygon} & \revised{45} & \revised{83} & \revised{84}  & \revised{84}  \\
\bottomrule
\end{tabular}
\label{tab:eval:capability}
\end{table}
 
\noindent
\textbf{Capability of identifying non-equivalent queries.}
We evaluate the capability of \sys, \verieql, and \polygon to identify non-equivalent queries.
Specifically, we vary the bound from 1 to 4 and check how many test cases each verifier can prove non-equivalent at the same bound.
The maximal bound is 4 because the number of non-equivalent cases identified by each verifier does not change when we continue to increase the bound.
To fully reflect the verification capabilities of all three verifiers, we wait until each verifier finishes normally without timing out.
\revised{
As shown in \Cref{tab:eval:capability}, on both the Spider and \spark benchmarks, \sys disproves more non-equivalent test cases than \verieql and \polygon.
}
\verieql disproves fewer non-equivalent cases than \polygon at all bounds, confirming the stronger verification capability of \polygon reported by Zhao et al.~\cite{zhao2025polygon}.
Compared with \polygon, \sys disproves 86, 115, 125, and 129 more non-equivalent test cases in the Spider benchmark at bounds 1, 2, 3, and 4, respectively.
\revised{
On the \spark benchmark, \sys disproves 18, 7, 26, and 26 more non-equivalent test cases at bounds 1, 2, 3, and 4, respectively.
}
Among 271 non-equivalent cases in the Spider benchmark, \sys can disprove 269 (99.3\%) of them at bound 4, while \polygon can disprove only 140 (51.7\%).
We find that the remaining two cases can be disproved at bound 10.
Among all 127 cases in the \spark benchmark, \sys can disprove 110 (86.6\%) of them at bound 4, while \polygon can disprove only 84 (66.1\%).
\revised{
The remaining 17 non-equivalent cases cannot be disproved by \sys because of unsupported SQL features, such as window functions.
}
The main reason for the advantage of \sys is that \sys can support more SQL features than \polygon, as discussed above.
Most cases are due to this reason.
Another reason is that for some non-equivalent SQL queries, \sys requires a smaller bound than \polygon to disprove them.
For the \cubes benchmark, since it uses only SQL features supported by \sys, \verieql, and \polygon, we mainly use it to evaluate verification speed, as presented in \Cref{subsec:eval:speed}.

\vspace{2pt}
\noindent
\textbf{Capability of supporting SQL features.}
We also demonstrate that \sys supports more SQL features by comparing the numbers of test cases supported by \sys and \polygon; \polygon has stronger verification capability than \verieql.
\revised{
\sys supports all SQL features supported by \polygon, including \sel, \proj, CTEs, subqueries, aggregate functions, \union, \unionall, \values, \innerj, \crossj, \fullj, \outerj, \order, \limit, \case, and \exist.
Some features are unsupported by \polygon but supported by \sys, such as \intersect, \excepts, and \off.
}
Specifically, we set the bound to one and run each verifier until it finishes normally.
If the verifier finishes normally on a test case, it means that the verifier can support it.
\revised{
Among 1,124 test cases in the Spider and \spark benchmarks, \sys supports 1,107, which is 337 more than \polygon supports.
}
Some cases unsupported by \polygon use SQL features, such as \intersect and \excepts.
In contrast, these SQL features can be directly encoded with U-expressions in \sys.
The result demonstrates the expressiveness of U-expressions for bag semantics.
Note that although \revised{U-expressions} are very powerful, they still cannot realize a complete verifier without our decision procedure (\Cref{sec:approach}).
\revised{
Besides, there are 178 cases using \order/\limit/\off, which cannot be supported by a U-expression alone.
}
\sys can support them due to the encoding method for ordered bag semantics (\Cref{sec:vc}).
The \cubes benchmark was crafted by \polygon and uses only SQL features \revised{supported by \polygon}.
Thus, \sys supports all features in the \cubes benchmark.

\subsection{Verification Efficiency}
\label{subsec:eval:speed}

Since \verieql can disprove only a small proportion of non-equivalent test cases, we mainly compare the verification time of \sys and \polygon to demonstrate the efficiency advantage of \sys.
We only consider the test cases supported by both \sys and \polygon for fairness.
Among these cases, we consider only non-equivalent test cases because complete verifiers output ``equivalent'' by default when they cannot finish within the predefined time limit. Thus, verification efficiency is meaningful only for non-equivalent cases.
For test cases from Spider and \spark, we increase the bound from 1 until the verifier can disprove it or the bound reaches the maximal value, 4, which is the same as the setting of \Cref{subsec:eval:capability}.
For test cases from \cubes, we increase the bound from 1 to 3 because neither verifier identifies new non-equivalent test cases when the bound is increased from 2 to 3.

\begin{table}[t]
	\caption{The complexity of benchmarks and median verification time (ms) of \sys and \polygon. ``\# AST Nodes of SQL'' is the sum of AST nodes of all queries in a test case. For fairness, we only compare the verification time of the test cases that are supported by both verifiers and contains non-equivalent queries.}
	\label{tab:eval:speed}
			\begin{tabular}{lrrrrrcc}
			\toprule
\multirow{2}{*}{\textbf{}} & \multicolumn{3}{c}{\textbf{\# AST Nodes of SQL}} & \multirow{2}{*}{\textbf{\sys}} & \multirow{2}{*}{\textbf{\polygon}}\\
\cmidrule(l){2-4} 
 & Avg. & Med. & Max.  &  &  \\
\midrule
Spider & 43   & 36  & 158  & 8.0 &  47.4\\
\revised{\spark} & \revised{46} & \revised{41} & \revised{98} & \revised{12.0} & \revised{53.5} \\
\cubes & 4251 & 4219 & 7233 &  95.0 & 13426.3 \\
\bottomrule
\end{tabular}
\end{table}

\Cref{tab:eval:speed} compares verification efficiency and test-case complexity in terms of the number of AST nodes.
Specifically, for each test case, we parse all its queries with the popular SQL parser SQLGlot~\cite{sqlglot} and count the nodes in the generated AST.
\sys brings more performance benefits on the \cubes benchmark.
The median verification time is reduced from 13.4 seconds to only 95 ms, a speedup of more than 100$\times$.
This is because the test cases in the \cubes benchmark are much more complex.
The median number of AST nodes in the \cubes benchmark is $>100\times$ that in the other two benchmarks.

\begin{figure}[t]
\centering
\subfloat[\revised{Spider}]{%
\label{fig:eval:spider}%
\begin{minipage}{0.32\linewidth}
\centering
\includegraphics[width=\linewidth]{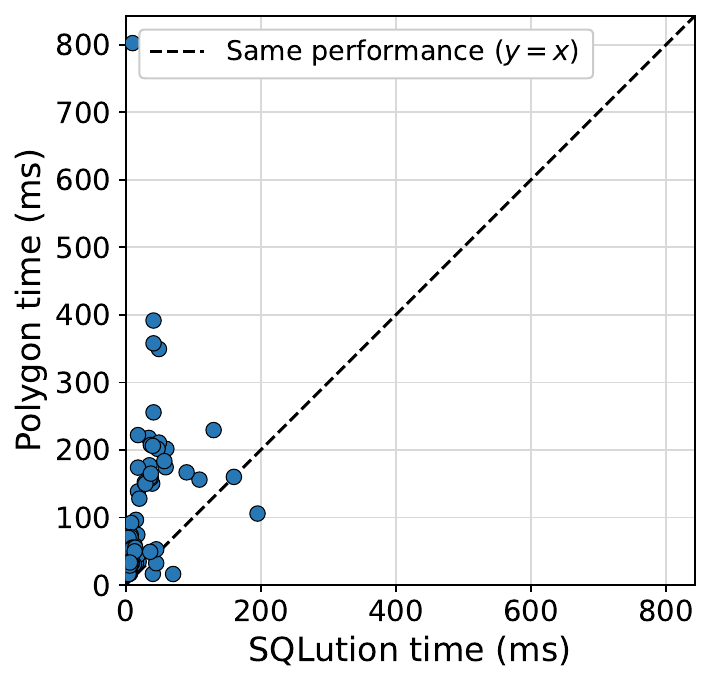}
\end{minipage}%
}\hfill%
\subfloat[\revised{\spark}]{%
\label{fig:eval:spark}%
\begin{minipage}{0.32\linewidth}
\centering
\includegraphics[width=\linewidth]{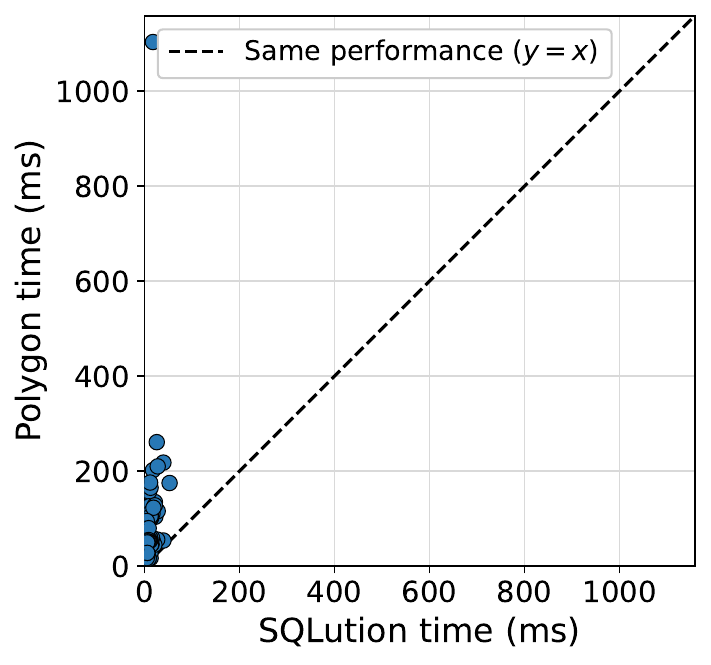}
\end{minipage}%
}\hfill%
\subfloat[\revised{\cubes}]{%
\label{fig:eval:cubes}%
\begin{minipage}{0.32\linewidth}
\centering
\includegraphics[width=\linewidth]{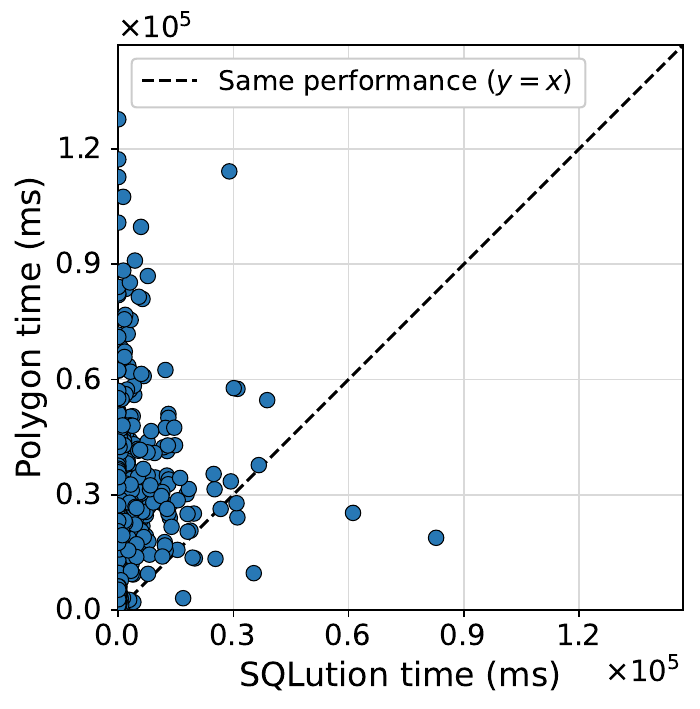}
\end{minipage}%
}
\caption{\revised{Per-case verification time of \sys and \polygon on the Spider, \spark, and \cubes benchmarks.
Each point represents the verification times for one test case shown to be non-equivalent by both verifiers. The dashed lines indicate equal performance.}}
\label{fig:eval:cdf}
\end{figure}

\begin{table*}[t]
\caption{Reduction of formula complexity of \sys over \polygon. For each bound value, we collect test cases that: (1) both verifiers can disprove it, and (2) it cannot be disproved with a smaller bound. We report the median reduction of complexity.}
\label{tab:eval:formula_complexity}
\centering
\small
\setlength{\tabcolsep}{4pt}
\begin{tabular}{crrrrrrrrrr}
\toprule
\multirow{2}{*}{\textbf{Bound}}
& \multicolumn{3}{c}{\textbf{Spider}}
& \multicolumn{3}{c}{\revised{\textbf{\spark}}}
& \multicolumn{3}{c}{\textbf{\cubes}} \\
\cmidrule(lr){2-4} \cmidrule(lr){5-7} \cmidrule(lr){8-10}
& \textbf{Size} & \textbf{\# AST Nodes} & \textbf{\# Operators}
& \revised{\textbf{Size}} & \revised{\textbf{\# AST Nodes}} & \revised{\textbf{\# Operators}}
& \textbf{Size} & \textbf{\# AST Nodes} & \textbf{\# Operators} \\
\midrule
1 & -53.0\% & -82.8\% & -83.1\%
  & \revised{-72.0\%} & \revised{-88.9\%} & \revised{-88.9\%}
  & -97.0\% & -98.7\% & -98.4\% \\
2 & -87.6\% & -96.0\% & -95.0\%
  & \revised{-82.1\%} & \revised{-93.8\%} & \revised{-92.6\%}
  & -98.8\% & -99.6\% & -99.4\% \\
3 & -81.6\% & -94.2\% & -93.7\%
  & \revised{-80.3\%} & \revised{-93.3\%} & \revised{-91.8\%}
  & / & / & / \\
\midrule
Med. & -59.2\% & -85.6\% & -86.0\%
     & \revised{-76.8\%} & \revised{-91.5\%} & \revised{-90.4\%}
     & -97.7\% & -99.1\% & -98.8\% \\
\bottomrule
\end{tabular}
\end{table*}
 
\revised{
\Cref{fig:eval:cdf} compares the verification times of all cases disproved by both \sys and \polygon.
Each point represents one test case.
The x-axis shows the verification time of \sys, while the y-axis shows that of \polygon. The $y=x$ line indicates equal performance.
Points above the dashed $y=x$ lines indicate cases in which \sys is faster than \polygon.
}
For the non-equivalent cases in the Spider benchmark, \sys reduces the verification time of 92.9\% of test cases by $\ge50$\%.
The verification time of 43.6\% of test cases can be reduced by $\ge80$\%.
\revised{
For the \spark benchmark, \sys reduces the verification time of 95.2\% of test cases by $\ge50$\%, and reduces the verification time of 51.2\% of test cases by $\ge80$\%.
}
For the \cubes benchmark, \sys reduces the verification time of 94.1\% of test cases by $\ge50$\%.
The verification time of 82.3\% of test cases can be reduced by $\ge90$\%.
\revised{
The median time reduction ratios of \sys over \polygon are 78.7\%, 82.1\%, and 97.2\% on Spider, \spark, and \cubes, respectively.
Among all test cases that can be disproved by both \sys and \polygon, \sys is faster than \polygon in 1,124/1,140 cases (98.6\%).
\sys is faster when 1) \sys generates a simpler SMT formula than \polygon, and 2) solving a simpler formula is faster for SMT solvers.
These two conditions do not always hold. The first condition may fail when the SQL queries are very simple and the number of table columns and the bound are very small.
The second condition does not always hold because of SMT instability~\cite{zhou2023mariposa}. It means that even two semantically equivalent SMT queries may sometimes have rather different performance on the same SMT solver.
}

We further analyze the reason for the efficiency advantage of \sys over \polygon.
We use three metrics to measure the complexity of a formula, including size, the number of AST nodes, and the number of operators.
The size is the length of the formula string in the SMT-LIB format~\cite{smtlib}.
For each bound, we consider only non-equivalent test cases satisfying two conditions: 1) both \sys and \polygon can disprove the case at this bound; and 2) neither verifier can disprove it at a smaller bound.
Thus, for the Spider and \spark benchmarks, we do not consider bound 4 because \polygon does not disprove new non-equivalent cases when the bound increases from 3 to 4.
For the \cubes benchmark, we do not consider bound 3 because neither verifier disproves new non-equivalent cases at this bound.
As shown in \Cref{tab:eval:formula_complexity}, \sys significantly reduces the formula size, the number of AST nodes, and the number of operators in all three benchmarks.
\revised{
For example, the median reductions in formula size, the number of AST nodes, and the number of operators are 76.8\%, 91.5\%, and 90.4\% on the \spark benchmark, respectively.
}
The reduction ratio is larger for the \cubes benchmark because test cases of the \cubes benchmark are more complex.
For the Spider and \spark benchmarks, the reduction ratio decreases when the bound increases from 2 to 3 because the test cases at different bounds are different.
Thus, \sys's simpler formulas explain its efficiency advantage.
\section{Related Work}
\label{sec:related}

\textbf{Complete verification for SQL equivalence.}
Recent works~\cite{chu2017cosette,wang2018oopsla,he2024verieql,zhao2025polygon} have made significant progress in developing complete verifiers for SQL query equivalence.
\verieql~\cite{he2024verieql} supports SQL features by representing the input and output of each SQL operator as symbolic lists related by FOL formulas.
This list-based encoding produces complex formulas, as discussed in \Cref{sec:back}.
\polygon~\cite{zhao2025polygon} accelerates verification by constructing under-approximations of \verieql's formulas to search a smaller space, but retains the complexity of list-based encoding.
In contrast, \sys uses U-expressions to generate much simpler formulas while supporting more SQL features.

\noindent
\textbf{Reasoning about U-expressions.}
U-expressions are an algebra capable of encoding a wide range of SQL features.
Prior works~\cite{chu2018udp,wang2024qed,wang2022wtune,sigmod2024sqlsolver} propose several algorithms for reasoning about their equivalence.
\udp~\cite{chu2018udp} uses predefined rules to normalize U-expressions and compares the resulting syntactic structures, but equivalent expressions do not always have structurally similar normal forms.
\wetune~\cite{wang2022wtune} instead translates equivalence into a FOL formula checked by SMT solvers, using manually crafted approximations that handle only specific forms of unbounded summation.
\sqlsolver~\cite{sigmod2024sqlsolver} handles arbitrary forms of unbounded summation without predefined rules.
Because all these approaches target soundness, \sys requires different algorithms to ensure completeness.

\noindent
\textbf{Reasoning about ordered bag semantics.}
Some existing verifiers~\cite{wang2022wtune,zhou2019equitas,zhou2022spes,wang2024qed,chu2018udp} only support bag semantics.
Recently, some works~\cite{sigmod2024sqlsolver,he2024verieql,zhao2025polygon} have proposed methods to support queries with ordered bag semantics.
Existing methods are incomplete or inefficient; \sys is complete and faster.
\section{Limitations}
\label{sec:discussion}

\revised{
\sys has the following two limitations.
First, \sys does not support some SQL features, such as window functions, \sqlover, \partitionby, \textsf{ROLLUP}, \textsf{CUBE}, lateral scalar sub-query, and recursive CTEs, because these features are not supported by U-expressions.
They are also unsupported by the evaluated complete verifiers \verieql and \polygon~\cite{he2024verieql,zhao2025polygon}.
Second, if a query sorts tuples by non-unique attributes, its result will be nondeterministic because the order of tuples with identical values for the sorting attributes is ambiguous.
Existing SQL equivalence verifiers~\cite{sigmod2024sqlsolver,he2024verieql,zhao2025polygon} do not define query equivalence in such cases.
Thus, \sys also does not handle such ambiguous queries.
How to encode more SQL features into U-expressions and handle nondeterministic SQL queries is an interesting direction for future work.
}
\section{Conclusion}
\label{sec:conclusion}

\revised{
This paper presents \sys, a complete verifier for SQL query equivalence.
We propose a new encoding method for SQL query equivalence based on U-expressions that produces much simpler formulas than prior work.
We design a complete decision procedure to reason about U-expressions, which can handle unbounded summations with unbounded summation expansion.
The evaluation shows that \sys can verify queries dramatically faster than the state-of-the-art \polygon while supporting more SQL features.
}

\bibliographystyle{ACM-Reference-Format}

\end{document}